\documentclass[prd,aps,preprintnumbers,nofootinbib,showpacs,superscriptaddress,amsfonts,floatfix]{revtex4}
\usepackage{epsfig}
\usepackage{graphicx}

\newcommand{\be}{\begin{equation}}

\newcommand{\ee}{\end{equation}}
\newcommand{\bea}{\begin{eqnarray}}
\newcommand{\eea}{\end{eqnarray}}

\def\starteq{%\vspace{.1in}
\begin{eqnarray}}
\def\fineq{\end{eqnarray}}

\def\a{\alpha}
\def\b{\beta}
\def\d{\delta}      
\def\e{\epsilon}

\def\g{\gamma}      \def\G{\Gamma}

\def\l{\lambda}   
\def\m{\mu}
\def\n{\nu}
   
\def\p{\pi}      
\def\q{\theta}   
\def\r{\rho}

\def\rarr{\rightarrow}

\begin{document}
\preprint{BARI-TH 516/05}
\preprint{NAPOLI DSF-2005/33}
\title{Role of final state interactions in the $B$ meson decay into two pions}
\author{Aldo Deandrea}
\affiliation{Universit\'e de Lyon 1, Institut de Physique Nucl\'eaire, Villeurbanne Cedex, France}
\author{Massimo Ladisa}
\affiliation{Dipartimento di Fisica dell'Universit{\`a} di Bari, Italy}
\affiliation{Istituto Nazionale di Fisica Nucleare, Sezione di Bari, Italy}
\author{Vincenzo Laporta}
\affiliation{Dipartimento di Fisica dell'Universit{\`a} di Bari, Italy}
\affiliation{Istituto Nazionale di Fisica Nucleare, Sezione di Bari, Italy}
\affiliation{Universit\'e de Lyon 1, Institut de Physique Nucl\'eaire, Villeurbanne Cedex, France}
\author{Giuseppe Nardulli}
\affiliation{Dipartimento di Fisica dell'Universit{\`a} di Bari, Italy}
\affiliation{Istituto Nazionale di Fisica Nucleare, Sezione di Bari, Italy}
\author{Pietro Santorelli}
\affiliation{Dipartimento di Scienze Fisiche, Universit{\`a} di Napoli ``Federico II'', Italy}
\affiliation{Istituto Nazionale di Fisica Nucleare, Sezione di Napoli, Italy}
\begin{abstract}
\noindent
We estimate final state interactions in the $B$-meson decays into
two pions by the Regge model. We consider Pomeron exchange and the
leading  Regge trajectories that can relate intermediate particles
to the final state. In some cases, most notably $B\to \p^0\p^0$ and
$B\to \p^+\p^-$, the effect is relevant and produces a better agreement
between theory and experiment.
\end{abstract}
\maketitle

\section{Introduction \label{sec:-1}}

Measuring the angle $\alpha$ of the unitarity triangle is one of
the major challenges of the  $B$-factories BaBar  at SLAC,  BELLE
at KEK and the future LHC at CERN. $B\to\pi\pi$ decay channels
were identified long ago as a promising candidate for the
extraction of the angle $\alpha$. Though other channels were
subsequently investigated, the $B$ decay into two pions is still
object of intense studies, both experimental and theoretical. The
task of determining precisely the angle $\alpha$ is complicated by
the problem of disentangling two different hadronic matrix
elements, each one carrying its own weak phase. They  are usually
referred to as the tree and penguin contributions.

The theoretical uncertainty in the evaluation of these terms stems
from the approximate schemes used to compute the relevant
four-quark operators between the hadronic states. In principle one
could avoid these uncertainties, fixing all the hadronic
parameters by a simultaneous measurement of physical observables.
This problem has been addressed by several authors; for example in
Ref. \cite{Gronau:1990ka} the use of isospin symmetry among the
various $B\to\p\p$ channels was envisaged. This program represents
a significant experimental challenge and is therefore useful to
have some theoretical indications on the results. In the Standard
Model one expects the dominance of tree diagrams in $B \to
\pi\pi$ decays, differently from the $B \to \pi K$ decay channels,
where  penguin contributions should play a key-role. Thus one
naively expects that  for $B \to \pi\pi$ the hierarchical
structure follows the analogous hierarchy of the Wilson
coefficients, namely
\begin{equation}
\label{eq:hierarchy}
 {\mathcal B}(\pi^0\pi^0) \ll {\mathcal B}(\pi^+\pi^-)
  \approx 2\ {\mathcal B}(\pi^+\pi^0) \ .
\end{equation}
As discussed in section \ref{sec:2} below, present experimental
data are at odds with Eq. (\ref{eq:hierarchy}).
There can be several factors leading to
violations of the expectation (\ref{eq:hierarchy}). First of all
the role of penguin operators should not be neglected. Second,
one has to go beyond naive factorization and use more
sophisticated schemes taking into account QCD in factorization,
for example the BBNS approach \cite{Beneke:1999br,Beneke:2000ry}
or the Soft-Collinear-Effective-Theory (SCET) \cite{Bauer:2000ew,Bauer:2000yr,Bauer:2001ct,Bauer:2001yt}
(for a recent discussion of $B$ decay into two light mesons in the framework of the SCET
see  \cite{Bauer:2004tj}). Finally, (\ref{eq:hierarchy})
does not take into account  final state interactions (FSI). To this issue the present
paper is devoted.
FSI are long distance effects that in some cases might play a significant role; for example in  $B$ decays
into two light  mesons a source of long-distance contributions is
provided by the  charming penguin diagrams that might produce the
discrepancy between experimental data and the naive factorization
findings. Charming penguins are contributions where the final
state is formed only as an effect of a rescattering process, and
is preceded by the formation of an intermediate
 state containing a $c\bar c$ pair
\cite{Colangelo:1989gi,Ciuchini:1997hb,Isola:2001ar,Isola:2001bn,Isola:2003fh,Colangelo:2004rd}.
In particular for decay channels with a strange light meson in the
final state, e.g. $B\to\p K$, these long-distance contributions
are not numerically suppressed. In fact, the
Cabibbo--Kobayashi--Maskawa (CKM) matrix elements produce  an
enhancement $\sim |V_{cb}V^*_{cs}|/|V_{ub}V^*_{us}|$ which can
compensate the parametric suppression predicted by QCD
factorization.

The role of charming penguins in $B\to\p\p$ is less clear. Due to
the lack of the above-mentioned enhancement, on general ground one
expects a minor role in the $B \to \pi\pi$ decay modes. On the
other hand in \cite{Bauer:2004tj} their role is found to be
significant. This matter should be settled, but in any case  FSI
must be taken into account, be they dominated by charming penguins
or by other rescattering processes, involving non-charmed
particles in the intermediate state.

The most accurate way to take into account FSI in hadronic $B$
decays is provided by the Regge model of high energy scattering
processes, which can be applied to hadronic $B$ decays due to the
rather large value of $s=m^2_B$. The advantage of the Regge
approach is to evaluate the rescattering  not by a Feynman
diagram, but by unitarity diagrams and the Watson's theorem
\cite{Watson:1952ji}. In particular there is no extrapolation of
low energy effective theories to the  hard momenta regime and
therefore no need to introduce arbitrary cutoffs in the light
meson momenta, because the high energy behavior is completely
under control. Some studies on the application of the Regge model
to $B$ decays are in Refs.
\cite{Zheng:1995mg,Donoghue:1996hz,Nardulli:1996ht}. Elastic
contributions to high energy scattering are dominated by the
Pomeron exchange, while the inelastic channels get contributions
from both Pomeron and Regge trajectories. Also charming penguins
find a place in this scheme, provided one introduces also charmed
Regge trajectories, as for example in the study performed in
\cite{Ladisa:2004bp} for the charmless $B$ decay into two light
vector mesons.

The aim of this letter is to extend the results of the Regge model
to the $B \to \pi\pi$ decay modes. We will show that there are
indeed significant rescattering
effects in the $B\to\p^0\p^0$ channel, a decay mode that is
suppressed in naive factorization. We include several intermediate
states: $\p\p$, $\r\r$, $a_1\p$, $D\bar D$ and we find that the
largest contribution comes from the $\p\p$ and  $\r\r$
intermediate states with $\r$ and $a_2$ Regge exchanges
respectively. On the other hand we find no significant role of
charming penguin contributions. The suppression of charming
penguins in comparison to other terms is produced because Regge
charmed trajectories have a negative intercept $\alpha(0)$ and
therefore a suppression factor $(s/s_0)^{\alpha(0)}$ ($s_0\simeq
1$ GeV$^2$, a threshold).

The plan of the paper is as follows. In section \ref{sec:0} we
evaluate in the factorization approximation the bare amplitudes,
including tree and penguin contributions with no FSI (in
particular no charming penguins). In section \ref{sec:1} we
discuss rescattering effects parameterized by the Regge model.
Finally, in section \ref{sec:2} we present our numerical results
and discuss them.
\section{Bare amplitudes \label{sec:0}}
The $\p\p$ final state can be reached from several intermediate
states {\it via} rescattering.  Clearly one should select  the
most prominent channels. Among the inelastic channels we single
out the decays $B\to\r\r$ and $B\to a_1\p$ since they have large
branching ratios. For example: ${\cal B}(B^+\to\r^0\r^+)=(26.4\pm
6.1)\times 10^{-6}$; ${\cal B}(B^0\to\r^-\r^+)=(30.0\pm 6.0)\times
10^{-6}$; ${\cal B}(B^0\to a_1^-\p^+)=(42.6\pm 5.9)\times 10^{-6}$
\cite{Alexander:2005cx}. To these decay processes we have to add
the elastic $B\to\p\p$ channels, though they have smaller
branching ratios. We also add the $D^{(*)}\bar D^{(*)}$, having in
mind a discussion on the charming penguins. In conclusion the
final state interactions that we consider are  the elastic
scattering $\p\p\to\p\p$, and the $\r\r\to\p\p$ and $a_1\p\to\p\p$
and  the $D^{(*)}\bar D^{(*)}\to\p\p$ inelastic channels.

We evaluate  bare amplitudes in the factorization approximation. To
do that one needs different input parameters. The non-leptonic
hamiltonian is well known and we do not repeat it here, see e.g.
\cite{Ali:1998eb}. For the Wilson coefficients we use: $a_1=1.029$,
$a_2=0.140$, and $(a_3,a_4,a_5,a_6,a_7,a_8,a_9,a_{10})$
$=(33.33,-246.66,-10,-300,1.95,4.81,-93.30,-12.63)$  $\times
10^{-4}$ \cite{Buras:1998ra}. We use for the  parameterization of
the CKM matrix $\sin\q_{12}=0.2243$, $\sin\q_{23}=0.0413$,
$\sin\q_{13}=0.0037$ and $\d_{13}=\g=1.05$ \cite{Eidelman:2004wy}.
As for the form factors and constant decay we use $f_\p=0.132$ GeV,
$f_\r=0.210$ GeV, $f_{a_1}\approx 0.21$ GeV (see the discussion in
\cite{Nardulli:2005fn}) $F_1^{B\to\p}(0)=0.26$,
$A_1^{B\to\r}(0)=0.26$, $A_2^{B\to\r}(0)=0.23$ , $V_0^{B\to
a_1}(0)=A_0^{B\to\r}(0)=0.39$, where we use the notations of
\cite{Ball:2003rd} for the $B\to\r$ transition and the
parameterization of Ref. \cite{Deandrea:1998ww} for the $B\to a_1\p$
matrix element. All the other parameters are taken from
\cite{Eidelman:2004wy}. We get in this way the results of Tables
\ref{tab0}-\ref{tab000} (notice that units of Tables
\ref{tab0}-\ref{tab00} are $10^{-8}$ GeV, those of \ref{tab000} are
$10^{-7}$ GeV).
%-----------------------------------------------------------------------
\begin{table}[ht!]
\begin{center}
\begin{tabular}{|c|c||c|c|c|c|}
\hline
\hspace{0.5truecm} Process  \hspace{0.5truecm} &
\hspace{0.8truecm} $A_b$    \hspace{0.8truecm}   &
\hspace{0.5truecm} Process  \hspace{0.5truecm} &
\hspace{0.3truecm} $A_b$ $(\l=+1)$ \hspace{0.3truecm} &
\hspace{0.3truecm} $A_b$ $(\l=-1)$ \hspace{0.3truecm} &
\hspace{0.3truecm} $A_b$ $(\l=0)$  \hspace{0.3truecm}
\\
\hline
$B^+\to\p^+\p^0$ & $+2.02-1.24\,i$ & $B^+\to\r^+\r^0$ &
$-0.02+0.01\,i$& $-1.1+0.65\,i$&
$+4.49-2.76\,i$
\\
$B^0\to\p^0\p^0$ & $-0.41+0.053\,i$ & $B^0\to\r^0\r^0$ &
$+0.004-0.001\,i$& $+0.20-0.07\,i$& $-0.83+0.31\,i$
\\
$B^0\to\p^+\p^-$ & $+2.43-1.74\,i$ & $B^0\to\r^+\r^-$ &
$-0.02+0.02\,i$& $-1.31+0.85\,i$& $+5.53-3.59\,i$
\\
\hline
\end{tabular}
\caption{Bare amplitudes for $B\to\p\p$ and $B\to\r\r$.
Results in $10^{-8}$ GeV; $\l=\pm 1,\,0$ refers to the helicities
of the vector particles. \label{tab0}}
\end{center}
\end{table}

\begin{table}[ht!]
\begin{center}
\begin{tabular}{|c|c|}
\hline
\hspace{0.5truecm} Process \hspace{0.5truecm} &
\hspace{0.8truecm} $A_b$ \hspace{0.8truecm}
\\
\hline
$B^+\to a_1^+\p^0$ & $+3.4-2.0\,i$
\\
$B^+\to\,a_1^0\p^+$ & $+2.2-1.6\,i$
\\
$B^0\to\,a_1^0\p^0$ & $-0.60+0.20\,i$
\\
$B^0\to\,a_1^+\p^-$ & $+4.2-2.7\,i$
\\
$B^0\to\,a_1^-\p^+$ & $+3.4-2.4\,i$
\\
\hline
\end{tabular}
\caption{Bare amplitudes for $B\to a_1\p$; $a_1$ with
longitudinal polarization. Units are $10^{-8}$ GeV. \label{tab00}}
\end{center}
\end{table}

\begin{table}[ht!]
\begin{center}
\begin{tabular}{|c|c||c|c|}
\hline
\hspace{0.7truecm} Process \hspace{0.7truecm} &
\hspace{0.4truecm} $A_b$ \hspace{0.4truecm} &
\hspace{0.7truecm} Process \hspace{0.7truecm} &
\hspace{0.4truecm} $A_b$ \hspace{0.4truecm}
\\
\hline $B^+\to D^+\bar D^0$ & $-2.8\,i$ &  $B^+\to D^{*+}\bar D^0$ &
$+2.5\,i$
\\
$B^0\to D^+ D^-$ & $-2.8\,i$ & $B^0\to D^{*+} D^-$  & $+2.5\,i$
\\
$B^+\to D^{*+}\bar D^{*0}$ & $-2.9\,i$&  $B^+\to D^+\bar D^{*0}$ &
$+2.5\,i$
\\
$B^0\to D^{*+} D^{*-}$ & $-2.9\,i$ & $B^0\to D^+ D^{*-}$  &
$+2.5\,i$
\\
\hline
\end{tabular}
\caption{Bare amplitudes $B\to D^{(*)}\bar D^{(*)}$; vector particles have
longitudinal polarization. Results in $10^{-7}$ GeV.
\label{tab000}}
\end{center}
\end{table}

\section{Final state interactions and Regge behavior \label{sec:1}}

Corrections to the bare amplitudes due to final state interactions are taken into
account by means of the Watson's theorem \cite{Watson:1952ji}:
\be
A = \sqrt{S}A_{b}
\label{amp}
\ee
where $S$ is the $S$-matrix,
$A_{b}$ and $A$ are the bare and the full amplitudes. An
application of the Watson's theorem was first discussed in
\cite{Donoghue:1996hz} and subsequently applied to other decay
channels in \cite{Nardulli:1996ht} and \cite{Ladisa:2004bp}.  We
briefly review here the formalism.

The two-body $S$-matrix elements are given by
\starteq
S^{(I)}_{ij} = \d_{ij} + 2i\sqrt{\r_{i}\r_{j}}\
T^{(I)}_{ij}(s)~~,
\label{s-matrix}
\fineq
\par\noindent
where $i,j$ run over all the channels involved in the final state
interactions. The $J=0$, isospin $I$ amplitude $ T^{(I)}_{ij}(s)$
is obtained by projecting the $J=0$ angular momentum out of the
amplitude $T^{(I)}_{ij}(s,t)$:
\starteq
T^{(I)}_{ij}(s) =  {1
\over 16\p}{s \over \sqrt{\ell_{i}\ell_{j}}}
\int^{t_{-}}_{t_{+}}dt \,T^{(I)}_{ij}(s,t) ~~.
\label{S-wave}
\fineq
$\r_j\,,\ell_j$ and $t_\pm$ are defined in Ref. \cite{Ladisa:2004bp}. For the channel $B\to \pi\pi$ we only have the
$I=0$ and $I=2$ transition amplitudes; the decay amplitude $B\to
\pi^+\pi^0$ is only $I=2$.

The phenomenological basis for the application of the Regge model
of final state interactions is the large value of $s=m^2_B$;
therefore a Regge approximation based on Pomeron exchange and the
first leading trajectories should be adequate. The Pomeron term
contributes to the elastic channels. As discussed in previous
section for the inelastic case we include only channels whose bare
amplitudes are prominent.

In conclusion in the present approximation we will include,
besides the Pomeron, the $\rho$ and $a_2$ (almost)
exchange-degenerate trajectories and $\pi$ Regge trajectories. We
shall discuss in subsection \ref{charm} the role of charmed Regge
trajectories in parameterizing charming penguins.

For the Pomeron contribution we write  (neglecting light meson masses)
\be
S \ =1+2iT^{{\cal P}}(s)\,,\hskip1cm T^{{\cal
P}}(s)=\frac{1}{16\p s }\int_{-s}^{0}T^{{\cal P}}(s,t) dt\, ,
\label{P0}
\ee
and we use the following parameterization \cite{Nardulli:1996ht,Donnachie:1992ny}:
\starteq
T^{{\cal P}}(s,t)=~-~ \b^{\cal P}g(t)
\left({s \over s_{0}} \right)^{\a_{P(t)}} e^{ -i\left(\p/2\right)\a_{P}(t)}~,
\label{P}
\fineq
with $s_{0} = 1\,\rm GeV^{2}$ and
$ \a_{P}(t) = 1.08 + 0.25t$ ($t {\rm \ in \ GeV^{2}})$, as given
by fits to hadron-hadron scattering total cross sections.  For
the Pomeron residue $\b^{\cal P}$ we assume factorization  with a
$t$-dependence given by \cite{Nardulli:1996ht,Donnachie:1992ny}
\starteq
g(t) = {1 \over (1-
t/m_{\r}^{2})^{2}} \simeq e^{2.8t}~~.
\label{gt}
\fineq
The additive quark counting rule allows to compute the Pomeron-pion
residue in terms of the Pomeron-nucleon ones. This gives
\cite{Zheng:1995mg,Nardulli:1996ht}:
\be\b^{\cal
P}_{\p}\sim \frac{2}{3} \b^{\cal P}_{p}\sim 5.1\ .
\ee
As observed in
\cite{Donoghue:1996hz} inelasticity effects play  an important
role in the determination of the FSI phases.
Parameterizing  them as in Ref. \cite{Donoghue:1996hz} by one effective state,
with no extra phases would allow to write the $S$-matrix as follows (neglecting a
small phase $\varphi=\,-0.01$ in $\sqrt{1+2iT^{{\cal P}}}$):
\be
(B\to \pi\pi)\hskip2cm  S\approx\left(%
\begin{array}{cc}
  0.62& 0.82\,i \\
 0.82\, i & 0.62 \\
\end{array}%
\right)\ , \hskip1.0cm \sqrt S\approx\left(%
\begin{array}{cc}
  0.79& 0.64(1+i) \\
  0.64(1+i) & 0.79 \\
\end{array}%
\right) \ .
\ee
This shows that even neglecting the effect of
the non leading Regge trajectories, final state interactions
due to inelastic effects parameterized by the Pomeron
exchange can produce sizeable strong phases.
This result agrees with the analogous findings of Refs.
\cite{Donoghue:1996hz}  and \cite{Nardulli:1996ht}. However this
method is not useful to evaluate rescattering effect in weak
decays. Therefore we prefer to parameterize inelastic effects by  Regge trajectories.

\subsection{Regge trajectories\label{reg}}

Let us now consider the contribution of the leading Regge
trajectories. Including Regge trajectories the  $S$ matrix can be
written for the generic  $B\to \pi\pi$ case as follows
\be
S = 1+ 2 i \left (T^{{\cal P}} +\sum {\cal R}\right )\, .
\label{regg}
\ee
Here ${\cal P}$ indicates the
Pomeron contribution discussed above. Since the Pomeron is much
larger than the others we make the approximation
\be
A(B \rarr\p\p)^{(I)}\approx \sqrt{1+2iT^{{\cal P}}} A^{(I)}_{b}\ +
\ \frac{1}{2\sqrt{1+2iT^{{\cal P}}}}
\sum_{k}\sum_{\cal R}
\left({\cal R}\right)^{(k,I)}
A_{b}^{(k)}\, .
\ee
Here the sum over $k$ refers to the various
intermediate states contributing to the final state $\p\p$; $I$ is
the isospin index.

We write the Regge amplitudes as follows
(${\cal R}=\r\,,a_2\,,\p$):
\be
{\cal R}^{(k,I)}(s)=\frac{1}{16\p\,s}
\int^{0}_{-s} dt {\cal R}^{(k,I)}(s,t)\, .
\ee
We assume the general parameterization
\starteq
{\cal R}^{(k,I)}(s,t)\approx -\b^{R}\, {1 +
(-)^{s_{R}} e^{ -i\p\a_{R}(t)} \over 2}\,\G(l_{R} - \a_{R}(t))\ (\a
')^{1 - l_{R}}\ (\a 's)^{\a _{R}(t)}
\label{R}
\fineq
as suggested in Ref. \cite{Irving:1977ea}. The trajectory is given by
\be
\alpha_R(t)=s_R+\alpha^\prime\,(t-m_R^2)\,=\,\alpha_R(0)\,+\,
\alpha^\prime\,t\ ,
\label{eq14}
\ee
with $\alpha^\prime=0.91$
GeV$^{-2}$. We notice the Regge poles at $l_{R} - \a_{R}(t)=0,-1,-2,\cdots$.
%------------------------------------------------------------
\begin{table}[ht!]
\begin{center}
\begin{tabular}{|c|c|c|c|}
\hline ~Trajectory ${\cal R}$~ & ~~$s_R$~~ & ~~$\ell_R$~~ &
~~~$\alpha_R(0)$~~~ \\
\hline $\rho$ & $1$ & $1$ &$ 0.5$ \\
$a_2$ & $2 $ & $1 $ & $0.5  $\\
$\pi$ & $ 0$ & $ 0$ & $\approx 0 $\\
\hline
\end{tabular}
\caption{Parameters of the Regge
trajectories. Exchange degeneracy is assumed.\label{tab1}}
\end{center}
\end{table}
The parameters we use are reported in Table \ref{tab1}. Near $t =
m_{R}^{2}$, Eq. (\ref{R}) reduces to
\starteq
{\cal R} \approx \b^{R}{s^{s_{R}}
\over ( t-m_{R}^{2} )}\ .
\label{Rapprox}
\fineq
We write $\beta^R=\beta_1^R\beta_2^R$ using factorization of the residues at
the two (1 and 2) vertices. Therefore Eq. (\ref{Rapprox}) allows to
identify $\b^{R} $ as the product of two on-shell coupling
constants. The residues  can be obtained by the decay rates
$\r\to\pi\pi$, $a_{1,2}\to\r\p$.  More precisely we obtain
$\beta^\rho_{\p^+\p^0}=8.2$ and $\beta^\rho_{a_1\p}\approx 2$ from
the $\rho\to\p\p$ and $a_1\to\r\p$ decay widths, respectively. Due to small value
of the residue $\beta^\rho_{a_1\p}$ we will neglect the contribution
of this channel in the sequel.

Let us now discuss the $a_2$ exchange. The residue $\beta^{a_2}_{\r^+\p^0}$ can be derived from
the strong coupling constant defined by
\be
{\cal M}(a_2^+(p,\eta)\to\r^+(k,\e)\p^0(q))=\frac{g_a}{m_{a_2}}\eta^{\m\n}
q_\m\e_{\n\a\beta\lambda}\e^{*\a}p^\beta q^\lambda\, .
\ee
From $a_{2}\to\r\p$ we get $g_a\approx 25$ GeV$^{-1}$. To compute the
residue we note that the $a_2$-exchange can only occur when the $\r$
intermediate particles have transverse polarization. Its residue is
related to $g_a$ by
$\beta^{a_2}_{\r^+\p^0}=g_a/\sqrt{\a^\prime}\approx 13.1$. This
phenomenological value is smaller than the theoretical value
given in \cite{Irving:1977ea}, on the basis of the Gell-Mann, Sharp
and Wagner model \cite{gellmann} for the $\omega\to3\p$ decay. In
view of the theoretical uncertainties arising from the hypothesis of
exchange-degeneracy and from the procedure we have described, we
will let this parameter vary with a spread of $\pm 50\%$ around a
central value, i.e. we assume
\be
\beta^{a_2}_{\r^+\p^0}=13.1\times(1\pm0.50) \ .
\label{range}
\ee
We note that, though the bare amplitudes $B\to\r\r$ with transversely
polarized $\r$'s are suppressed (see Table \ref{tab0}), they participate nevertheless in
the rescattering process due to the large residue of the $a_2$
trajectory to the $\r$ and $\p$.

As shown by table \ref{tab1} the
$\p$ trajectory is exponentally suppressed due to $\a_\p\approx 0$.
Similarly  we note that we have also computed the parameters of the
$a_1$ Regge pole, but we omit this trajectory from the analysis
because its intercept is large and negative ($\alpha(0)\approx
-0.37$).

\subsection{Charming penguins\label{charm}}

Charming penguins are diagrams describing the rescattering
of two charmed mesons to produce two light mesons. Treating them
as Feynman diagrams produces a huge theoretical uncertainty.
In fact to compute them one should employ the chiral effective
theory for light and heavy mesons. However this approach
cannot be extended to hard meson momenta and one is forced to
introduce a cut-off \cite{Isola:2001ar,Isola:2001bn,Isola:2003fh,Colangelo:2004rd}.
To avoid the arbitrariness of this procedure one can describe this
class of FSI by charmed Regge trajectories. This approach was
followed in \cite{Ladisa:2004bp} for $B\to\r\r,\,\
K^*\r,\,K^*\phi$ decays and can be easily extended to $B\to\p\p$;
we refer to this paper for details. Let us only write down the
expression of trajectories $\alpha_{D}(t)$ and $\alpha_{D^*}(t)$.
We use Eq. (\ref{eq14}) with $s_D=0$, $s_{D^*}=1$ and
\cite{Ladisa:2004bp}
\be
\alpha_0=-1.8\,,\hskip1cm
\alpha^\prime\,=\,(0.39\,\pm \,0.12) \ {\rm GeV} ^{-2}\ ,
\label{eq:27}
\ee
which shows that the intercept of these
trajectories is negative. Also the residues can be computed
following the procedure of previous subsection, i.e. using the
strong coupling constants $g_{D^*D\p}$  and $g_{D^*D^*\p}$ (for
the values of these constants we follow \cite{Isola:2001ar}).

Differently from the case of $K\pi$ or the $K^*\r$ final states,
the bare  $B\to D^{(*)}\bar D^{(*)}$ amplitudes have no CKM enhancement.
Therefore the situation is similar to the study of the
$B\to\r\r$ channel in  \cite{Ladisa:2004bp} where  we found that, for the $\r\r$
final state, charming penguins are less relevant than, for example,  $B\to
K^*\r$.  We have checked numerically that the negative intercept produces
a negligible contribution from the $D^{(*)}\bar D^{(*)}$ intermediate states to $B\to\p\p$.

\section{Numerical results and discussion\label{sec:2}}

\begin{figure}[ht]
\begin{center}
\centerline{
\epsfxsize=6truecm\epsffile{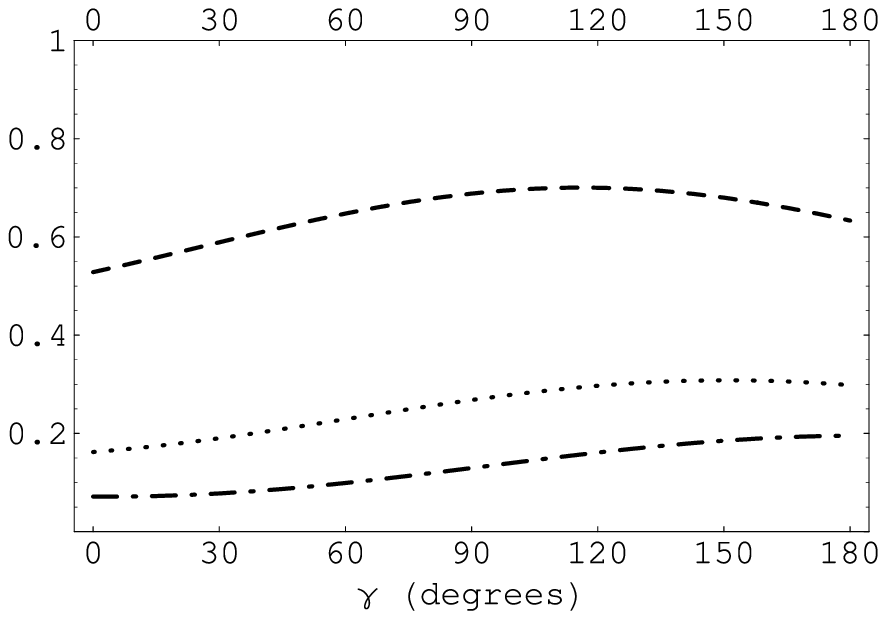}\hskip.2cm
\epsfxsize=6truecm\epsffile{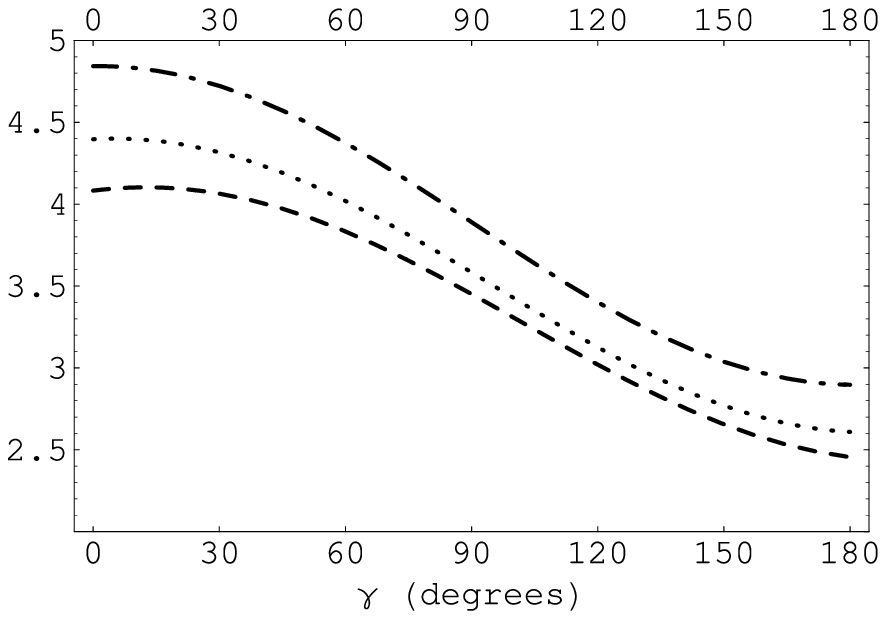}\hskip.2cm
\epsfxsize=6truecm\epsffile{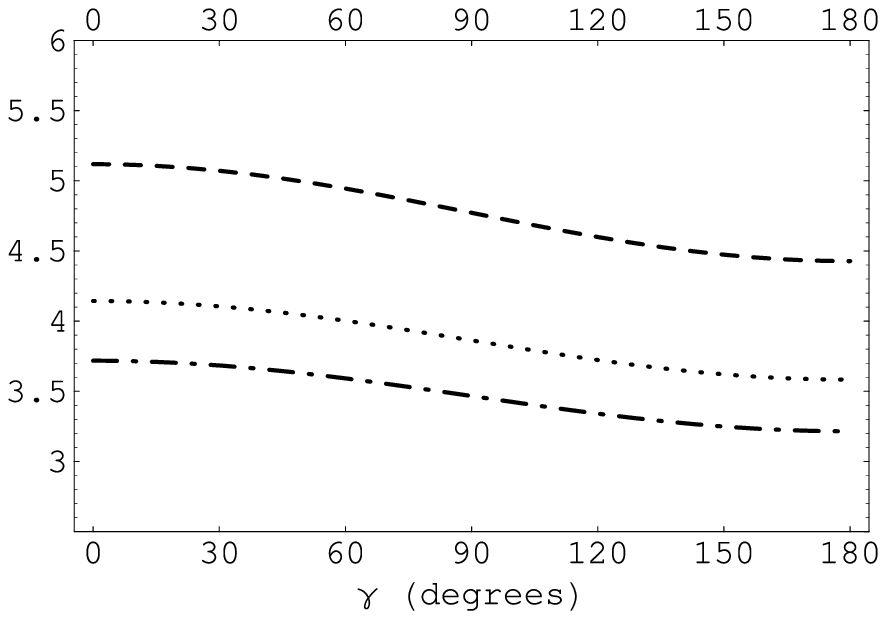}}
\noindent {\caption {\label{fig1}{Branching ratios (units
$10^{-6}$) for $B\to\p\p$ as functions of the angle $\g$ (degrees). From left
to right the decays $B^0\to\p^0\p^0$, $B^0\to\p^+\p^-$ and
$B^+\to\p^+\p^0$. Dashed,  dotted, and dot-dashed lines refer to the
upper, central and lower values of the Regge residue in Eq.
(\ref{range}).}}}
\end{center}
\end{figure}

We present our results by taking the $\g$ angle as a parameter and
allowing $\b^{a_2}_{\r^+\p^0}$ to vary in the range in Eq. (\ref{range}).
We compute in Fig. \ref{fig1} the branching ratios ${\cal
B}(B^0\to\p^0\p^0)$, ${\cal B}(B^0\to\p^+\p^-)$ and ${\cal
B}(B^+\to\p^+\p^0)$. A survey of the experimental results is in
table \ref{tab2}. Here we have also reported our results for
$\b^{a_2}_{\r^+\p^0}$ in the range of values given in Eq. (\ref{range}). We see
that the role of FSI is especially important for the $B\to\p^0\p^0$
channel.
\begin{table}[ht!]
\begin{center}
\begin{tabular}{|c|c|c||c|}
\hline
Process &
\hspace{0.2truecm} ${\cal B}$ (without FSI) \hspace{0.2truecm} &
\hspace{0.2truecm} ${\cal B}$ (with FSI) \hspace{0.2truecm} &
\hspace{0.2truecm} ${\cal B}$ (exp.) \hspace{0.2truecm}
\\
\hline
$B^0\to\p^0\p^0$ & $0.08$ & $0.10-0.65$ &
\begin{tabular}{cl}
$1.17\pm 0.32\pm 0.10$      &  \cite{Aubert:2002jb,Aubert:2004aq}\\
$2.3^{+0.4+0.2}_{-0.5-0.3}$ &  \cite{Chao:2003ue,Abe:2004mp}\\
$1.45\pm 0.29$              &  \cite{Alexander:2005cx}\\
\end{tabular}\\
\hline
$B^0\to\p^+\p^-$ & $8.1$ & $3.8-4.4$ &
\begin{tabular}{cl}
$4.7\pm 0.6\pm 0.2$ & \cite{Aubert:2002jb,Aubert:2004aq}\\
$4.4\pm 0.6\pm 0.3$ & \cite{Chao:2003ue,Abe:2004mp} \\
$4.5\pm 0.4$        & \cite{Alexander:2005cx}\\
\end{tabular}\\
\hline
$B^+\to\p^+\p^0$ & $5.0$ & $3.6-5.0$ &
\begin{tabular}{cl}
$5.8\pm 0.6\pm 0.4$ &  \cite{Aubert:2002jb,Aubert:2004aq} \\
$5.0\pm 1.2\pm 0.5$  & \cite{Chao:2003ue,Abe:2004mp}\\
$5.5\pm 0.6$        &  \cite{Alexander:2005cx}\\
\end{tabular}\\
\hline
\end{tabular}
\caption{Theoretical branching ratios for $B\to\p\p$ decay
channels with and without final state interactions and their
comparison with experimental data. The column FSI is computed with
$\b^{a_2}_{\rho^+\pi^0}$ in the range given in Eq. (\ref{range}).
Units $10^{-6}$. \label{tab2}}
\end{center}
\end{table}

\begin{figure}[ht]
\begin{center}
\centerline{
\epsfxsize=6truecm\epsffile{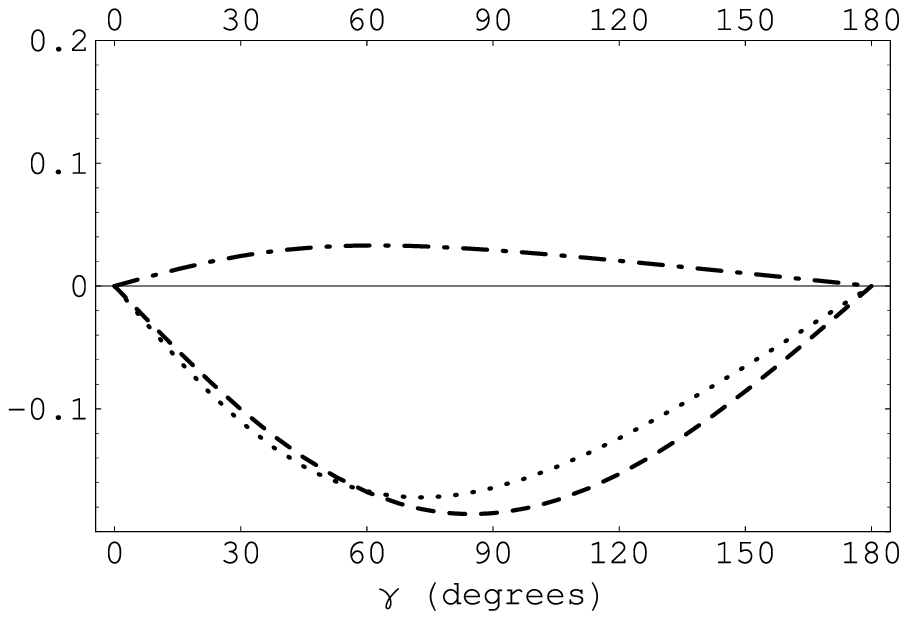} \hskip.0cm
\epsfxsize=6truecm\epsffile{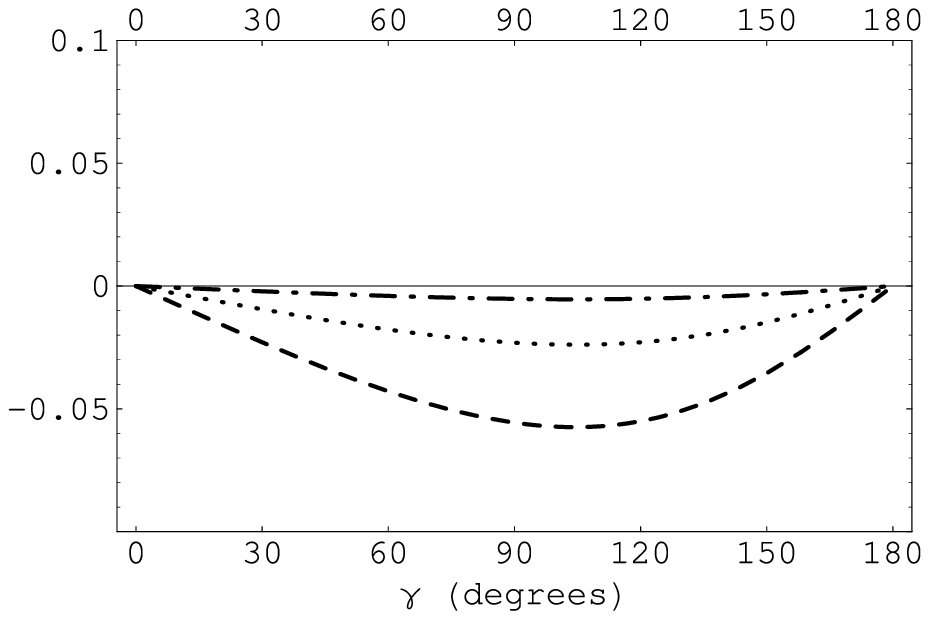} \hskip.0cm
\epsfxsize=6truecm\epsffile{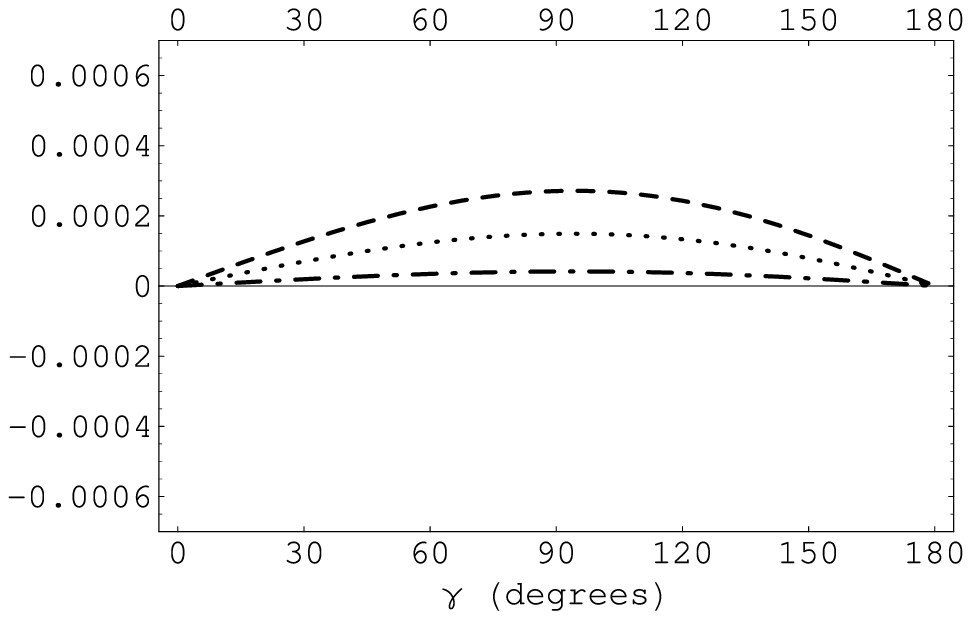}}
\noindent{\caption{\label{fig4}{Time integrated asymmetries
as defined in Eqns. (\protect\ref{asym}) as functions of the angle $\g$ (degrees).
From the left to right ${\cal A}_{00}$, ${\cal A}_{+-}$ and ${\cal
A}_{-0}$. Dashed, dotted, and dot-dashed lines refer to the upper,
central and lower values of the Regge residue in Eq. (\ref{range}).}}}
\end{center}
\end{figure}

Since we do not pretend to have presented a complete discussion of
final state interactions, our result should be interpreted as an
indication of the relevant role played by the rescattering effects
when the bare amplitudes are for some reason small. This is
confirmed by the results for the other two channels, where charged
current hamiltonian is involved and therefore  FSI play a less
relevant role. Nevertheless also for the
$B^0\to\p^+\p^-$ channel we can see that FSI contribution produce a
better agreement with the data,

We have also computed the integrated asymmetries
\bea
{\cal A}_{00}&=&\frac{\Gamma(\bar B^0\to\p^0\p^0)-\Gamma(
B^0\to\p^0\p^0)}{\Gamma(\bar B^0\to\p^0\p^0)+\Gamma(
B^0\to\p^0\p^0)} \ ,\nonumber\\
{\cal A}_{+-}&=&\frac{\Gamma(\bar
B^0\to\p^+\p^-)-\Gamma( B^0\to\p^+\p^-)}{\Gamma(\bar
B^0\to\p^+\p^-)+\Gamma( B^0\to\p^+\p^-)}\ ,\label{asym}\\
{\cal A}_{-0}&=&\frac{\Gamma(B^-\to\p^-\p^0)-\Gamma(
B^+\to\p^+\p^0)}{\Gamma(B^-\to\p^-\p^0)+\Gamma(
B^+\to\p^+\p^0)}\ .\nonumber
\eea
The results are reported in
Fig. \ref{fig4}. For ${\cal A}_{00}$ the HFAG group reports the
average of the BaBar and Belle Collaborations as follows
\cite{Alexander:2005cx}:  ${\cal A}_{00}=0.28\pm0.39$\,. For
$\g\simeq 60^\circ$ our result is compatible, within error with the
experiment.

Let us finally compare our results with other approaches. The exclusive
$B \to \pi\pi$ transitions can be studied, starting from first
principles, in the  {\it QCD factorization} approach (BBNS)
\cite{Beneke:1999br,Beneke:2000ry}. In this framework, all
the charmless two body decays of B mesons have amplitudes which are
shown to factorize at lowest order in $1/m_b$. In other words,
neglecting terms suppressed by heavy quark mass, the {\it QCD
factorization} predicts the {\it naive} factorization ansatz
\cite{Bauer:1986bm}. In this framework the branching ratios for the
$B \to \pi\pi$ decay modes are rather sensitive to the computational
scheme of the relevant form factor.
In particular, as discussed in \cite{Beneke:1999br}, results for
the $\p^0\p^0$ final state depend on a parameter $\lambda_b$ whose precise
value is unknown, but a branching ratio of the order of $10^{-6}$ could be reached.
In the phenomenological studies \cite{Beneke:2003zv,Cottingham:2005es}
of these processes, the authors take into account the power-suppressed
and partially unknown weak annihilation contributions. In particular,
the last fit to charmless strangeless final state alone in the BBNS
approach, performed in Ref. \cite{Cottingham:2005es}, reproduces
the available experimental data. Our method complements the
BBNS approach as it takes into account in a systematic
way part of the power suppressed contributions, i.e.
those arising from the final state interactions.

Agreement with experimental data on $B \to \pi\pi$ is also obtained
in \cite{Bauer:2004tj}, where the Soft Collinear Effective Theory
(SCET) \cite{Bauer:2000ew,Bauer:2000yr,Bauer:2001ct,Bauer:2001yt} is
employed. The BBNS and the SCET approaches substantially differ in
treating perturbative and non-perturbative effects, in particular
SCET predicts non negligible long-distance charming penguin
contributions. We do not discuss the differences between
QCD-factorization and the SCET as this goes beyond the limits of the
present work. We stress however that we do not find an important
role of the long distance charming penguin diagrams, but we find
another source of long distance effects due to rescattering of the
$\p\p$ and $\r\r$ channels. Charming penguins play a minor role here
because, as discussed in section \ref{sec:1}, in the Regge theory
they are strongly suppressed by the negative intercept of the
corresponding Regge trajectory. Our results are confirmed by a
different analysis \cite{Isola:2001bn} of the charming penguin
contributions in $B\to\p\p$, based on an effective lagrangian
approach; also in that paper these contributions play a lesser role
the reason being, there as in the present work, the absence of any
CKM enhancement in the bare amplitudes.

%\bibliographystyle{apsrev}
%\bibliography{Bphysics}

\end{document}